\documentclass{aa}

\usepackage{graphicx}
\usepackage{txfonts}

\begin{document}

\title{Buckling instability in tidally induced galactic bars}

\author{Ewa L. {\L}okas
}

\institute{Nicolaus Copernicus Astronomical Center, Polish Academy of Sciences,
Bartycka 18, 00-716 Warsaw, Poland\\\email{lokas@camk.edu.pl}}

\date{Received January 3, 2019; accepted February 18, 2019}

\abstract{
Strong galactic bars produced in simulations tend to undergo a period of buckling instability that weakens and thickens
them and forms a boxy/peanut structure in their central parts. This theoretical prediction has been confirmed by
identifying such morphologies in real galaxies. The nature and origin of this instability remains however poorly
understood with some studies claiming it to be due to fire-hose instability while others relating it to vertical
instability of stellar orbits supporting the bar. One of the channels for the formation of galactic bars is via the
interaction of disky galaxies with perturbers of significant mass. Tidally induced bars offer a unique possibility of
studying buckling instability because their formation can be controlled by changing the strength of the interaction
while keeping the initial structure of the galaxy the same. We use a set of four simulations of flyby interactions
where a galaxy on a prograde orbit forms a bar, which is stronger for stronger tidal forces. We study their buckling by
calculating different kinematic signatures, including profiles of the mean velocity in vertical direction, as
well as distortions of the bars out of the disk plane. Although our two strongest bars buckle most strongly, there is
no direct relation between the ratio of vertical to horizontal velocity dispersion and the bar's susceptibility to
buckling, as required by the fire-hose instability interpretation. While our weakest bar buckles, a stronger one does
not, its dispersion ratio remains low and it grows to become the strongest of all at the end of evolution.
Instead, we find that during buckling the resonance between the vertical and radial orbital frequencies becomes wide
and therefore able to modify stellar orbits over a significant range of radii. We conclude that the vertical orbital
instability is the more plausible explanation for the origin of buckling.}


\maketitle

\section{Introduction}

Bars are the dominant non-axisymmetric features in late-type galaxies and a significant fraction of galaxies of this
type possess them \citep[see e.g.][and references therein]{Buta2015}. Theoretical models have demonstrated that bars
can form as a result of instability in self-gravitating axisymmetric disks \citep[for a review
see][]{Athanassoula2013} or via interactions with perturbers of different masses \citep{Noguchi1996, Miwa1998}. Soon
after their formation stronger bars undergo one or more episodes of buckling instability that involves distortions of
the bar out of the equatorial plane of the disk and results in weakening and thickening of the bar \citep[for a review
see][]{Athanassoula2016}. The thickened bar takes the form of the so-called boxy/peanut shape when viewed edge-on.

Understanding buckling instability is important for the studies of bar evolution in galaxies, including the Milky Way,
which has been known for a long time to possess the boxy/peanut structure in the inner parts \citep{Weiland1994,
Ciambur2017}. Bar buckling in the Milky Way has been invoked to explain unusual behavior of stellar populations in the
bulge \citep{Debattista2017} and recently proposed by \citet{Khoperskov2019} to be the origin of the phase-space
spirals discovered in Gaia DR2 \citep{Antoja2018}. While the observations of bars confirmed the presence of boxy/peanut
shapes in many galaxies \citep{Bureau2006, Yoshino2015, Erwin2017, Li2017, Savchenko2017}, including the classical
example of NGC 128, catching the bars `in the act' of actual distortion out of the disk plane remains elusive.
\citet{Erwin2016} attempted this in the case of NGC 3227 and NGC 4569 by comparing asymmetric photometric and kinematic
features of these galaxies to those of a simulated buckling bar.

Buckling of galactic bars has been studied theoretically mostly via $N$-body and hydrodynamical simulations. The
presence of boxy/peanut morphology was first convincingly demonstrated in the simulations of \citet{Combes1981}.
\citet{Pfenniger1991} and \citet{Raha1991} detected periods of asymmetric distortion of their simulated bars out of the
disk plane before the formation of boxy/peanut shapes. \citet{Debattista2004} and \citet{Athanassoula2005} found that
the thickened parts of the bars resulting from buckling are in many aspects similar to bulges of late-type galaxies.
\citet{Martinez2006} were the first to report recurrent buckling in their simulation of a Milky Way-like galaxy. Later
simulations of \citet{Debattista2006} demonstrated that buckling is quite common in simulated galaxies and in general
weakens the bars but does not destroy them, contrary to the original suggestion of \citet{Raha1991}. Simulations
including gas dynamics and other hydrodynamical processes have however demonstrated that they can suppress buckling
\citep{Debattista2006, Berentzen2007, Villa2010}.

Although considerable effort has been invested into the studies of the buckling instability, its nature remains
unclear, as recently emphasized by \citet{Smirnov2018}. The early studies \citep{Combes1990, Pfenniger1991} suggested
that buckling is due to orbital instability resulting from coinciding horizontal and vertical inner Lindblad
resonances. The modification of the orbital structure then involves the bifurcation of the $x_1$ family of orbits into
$x_1v_1$ corresponding to the 2:1 vertical resonance and banana-like orbits, although the final orbital structure of
the boxy/peanut is much more complicated \citep{Patsis2002, Portail2015, Valluri2016, Abbott2017, Patsis2018}. Other
authors \citep{Raha1991, Merritt1991, Merritt1994} related buckling to the fire-hose instability as envisioned by
\cite{Toomre1966}. In this case the instability is controlled by the ratio of velocity dispersions $\sigma_z/\sigma_x$
along the vertical and horizontal directions. The instability is supposed to occur for low values of this ratio
(of the order of 0.3) and is prevented if the ratio is large enough. The discussion continues to this day with some
authors claiming the buckling to be due to orbital resonances \citep{Saha2018} while others assign it to the fire-hose
instability \citep{Zana2018}.

As mentioned above, one of the channels for the formation of bars is to induce them tidally by interaction with other
objects. These may result from tidal effects generated by a neighboring bigger structure, as in the case of a satellite
dwarf galaxy orbiting a bigger host \citep{Lokas2014, Lokas2015, Gajda2017, Gajda2018} or a normal-size galaxy orbiting
a cluster \citep{Mastropietro2005, Lokas2016}. Bars can also be induced by satellites infalling into or passing near a
bigger galaxy \citep{Mihos1995, Mayer2004, Gauthier2006, Purcell2011, Pettitt2018} or in
interactions between galaxies of similar mass \citep{Lang2014, Lokas2018, Peschken2019}. Tidally induced bars offer a
unique possibility of studying buckling instability because their formation can be controlled by changing the strength
of the interaction while keeping the initial structure of the galaxy the same. The bars forming in such interactions
are then of different strength but they share as many similarities as possible.

In this paper we study the buckling instability occurring in tidally induced bars formed in flyby interactions
of two Milky Way-like galaxies with the aim to elucidate its nature. In Section 2 we describe the simulations used
in this work and in Section 3 we characterize the bars formed in these simulations. Section 4 focuses on different
measures of the buckling instability and the discussion follows in Section 5.

\begin{table}
\begin{center}
\caption{Configuration details of the simulations.}
\begin{tabular}{lccccl}
\hline
\hline
Simulation  & $d$   &  $b$  &  $v$ &  $ S $ & Line color\\
            & (kpc) & (kpc) &(km s$^{-1}$)&  &\\
\hline
\ \ \ \ B1 &  500  & 25   & 500   &  0.07  & \ \ blue    \\

\ \ \ \ B2 &  350  & 25   & 350   &  0.15  & \ \ cyan   \\

\ \ \ \ B3 &  300  & 25   & 300   &  0.20  & \ \ green    \\

\ \ \ \ B4 &  250  & 25   & 250   &  0.26  & \ \ red     \\
\hline
\label{tabconfiguration}
\end{tabular}
\end{center}
\end{table}

\section{The simulations}

For the purpose of this study we used some of the simulations of galactic flybys described in \citet{Lokas2018} and
performed an additional one with a similar initial configuration. The configuration was illustrated in figure~1 of
\citet{Lokas2018} and comprised two Milky Way-like disky galaxies, each composed of a stellar disk and a dark matter
halo. The disks were aligned with the orbital plane ($XY$) of the flyby but so that one had exactly prograde and the
other exactly retrograde orientation. The galaxies were placed at a distance $d$ from the $Y$ axis and $b$ from the $X$
axis and assigned velocities $v$ in opposite directions along the $X$ axis of the simulation box. The distance $b$,
i.e. the nominal impact parameter, was the same for all simulations and the values of $d$ and $v$ were selected so that
the flyby took place approximately after 1 Gyr from the start of the simulation in all cases.

We considered four different choices of $d$ and $v$ listed in the second and fourth column of
Table~\ref{tabconfiguration} leading to four simulations which were labelled B1-B4. The fifth column of the Table gives
the values of the dimensionless tidal strength parameter $S$ of \citet{Elmegreen1991} which all fall in the interesting
regime of $S>0.04$ identified by \citet{Elmegreen1991} as leading to the formation of bars. The last column
indicates the color of lines with which the results for the corresponding simulation will be shown throughout the
paper. We note that the simulations B1, B2 and B4 were named S2, S3 and S4 in \citet{Lokas2018} and simulation B3 was
additionally performed for this study.

\begin{figure}
\centering
\includegraphics[width=9cm]{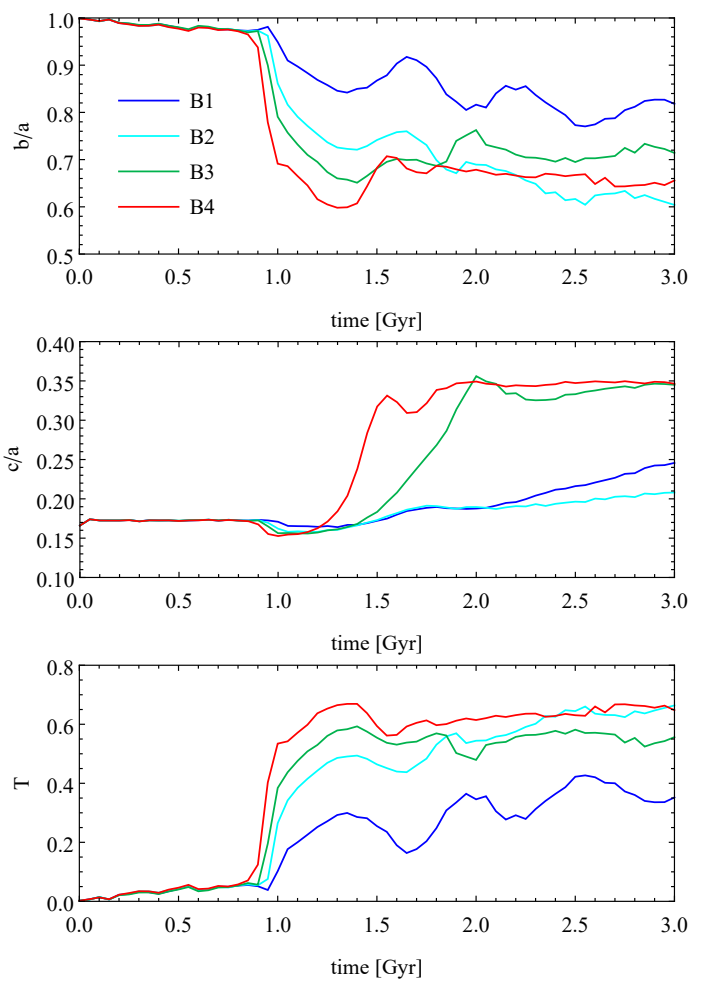}
\caption{Evolution of the shape of the stellar component in time. The three panels from top to bottom show respectively
the evolution of the axis ratio $b/a$ (intermediate to longest axis), $c/a$ (shortest to longest axis) and the
triaxiality parameter $T = [1-(b/a)^2]/[1-(c/a)^2]$. Measurements were made for stars within the radius of $2 R_{\rm
d}$.}
\label{shape}
\end{figure}

As discussed in \citet{Lokas2018}, the model of the Milky Way used here is stable against bar formation in isolation,
i.e. a very weak bar starts to form in the disk about 3 Gyr after the start of the simulation. For completeness, we
recall the structural parameters of the galaxy which was similar to the model MWb of \citet{Widrow2005}: its dark
matter halo had an NFW \citep{Navarro1997} profile with a virial mass $M_{\rm H} = 7.7 \times 10^{11}$ M$_{\odot}$ and
concentration $c=27$ while the exponential disk had a mass $M_{\rm D} = 3.4 \times 10^{10}$ M$_{\odot}$, the
scale-length $R_{\rm D} = 2.82$ kpc and thickness $z_{\rm D} = 0.44$ kpc. The profile of the Toomre parameter for this
model had a minimum of $Q=2.1$ at $2.5 R_{\rm D}$.

The model of the galaxy was same as the one used by \citet{Lokas2016}, \citet{Semczuk2017} and \citet{Lokas2018}. Its
$N$-body realization was created using the procedures described in \citet{Widrow2005}
and \citet{Widrow2008} with each component containing $10^6$ particles.
The evolution of the two galaxies was followed for 3 Gyr with the $N$-body code GADGET-2 \citep{Springel2001,
Springel2005} saving outputs every 0.05 Gyr. The adopted softening scales were $\epsilon_{\rm D} = 0.1$ kpc and
$\epsilon_{\rm H} = 0.7$ kpc for the disk and halo of the galaxies, respectively.

\section{Evolution of the bars}

As discussed in detail in \citet{Lokas2018}, out of the two galaxies participating in each flyby, only the one on the
prograde orbit forms a strong bar. The reasons for this behavior were thoroughly explained in this previous paper and
involve the resonance between the angular velocity of the stars in the prograde disk and the motion of the perturbing
galaxy. Here, from now on, we will consider only the bars formed in these prograde disks.

\begin{figure}
\centering
\includegraphics[width=9cm]{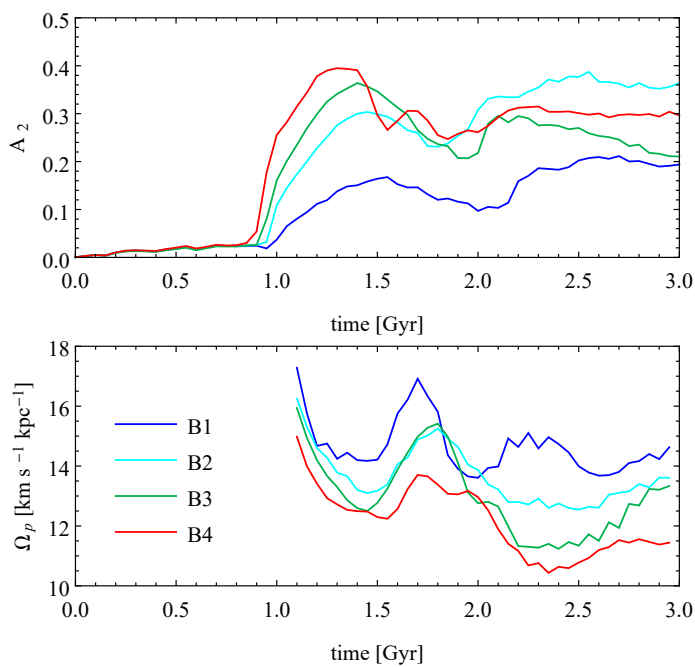}
\caption{Evolution of the bar mode amplitude $A_2$ (upper panel) and the pattern speed of the bar $\Omega_{\rm p}$
(lower panel) in time. Measurements were made for stars within the radius of $2 R_{\rm D}$.}
\label{barmodepatternspeed}
\end{figure}

The simplest and most straightforward way to describe the formation of the bar is to measure the evolution of the shape
of the stellar component of the galaxy in time. We do this by calculating the inertia tensor from all stars within
$2R_{\rm D}$ and aligning the stellar component with the resulting principal axes, so that $x$ is along the major axis,
$y$ along the intermediate axis and $z$ corresponds to the shortest one. After rotating the stellar component in this
way we calculate for each snapshot the axis ratios $b/a$ (intermediate to longest) and $c/a$ (shortest to longest)
using again all stars within $2R_{\rm D}$. The shape of the stellar component is immediately clear from the combination
of these ratios in the form of the triaxiality parameter $T = [1-(b/a)^2]/[1-(c/a)^2]$ which vanishes for infinitely
thin disks and approaches unity for thin needles.

The evolution of the three quantities is shown in Figure~\ref{shape}. At the beginning the properties of the galaxies
are very similar in all simulations and the disks are preserved, as demonstrated by the triaxiality parameter remaining
close to zero. At the time of the interaction, around 1 Gyr from the start of the simulations, abrupt changes take
place and the triaxiality increases very strongly reaching $T>0.5$, i.e. going from the low values characteristic of
disks to high values typical for prolate spheroids. This obviously signifies the formation of bars in all galaxies.
Note that right after the bar formation, around $t=1.3$ Gyr, higher values of triaxiality reflect the stronger
interactions during the flyby, as measured by the Elmegreen parameter (see Table~\ref{tabconfiguration}).

It is customary to measure the strength of the bar with the $m=2$ mode of the Fourier decomposition of the surface
distribution of stars projected along the short axis: $A_m (R) = | \Sigma_j \exp(i m \theta_j) |/N_s$ where $\theta_j$
is the azimuthal angle of the $j$th star and the sum is up to the total number of $N_s$ stars. The radius $R$ is the
standard radius in cylindrical coordinates in the plane of the disk, $R = (x^2 + y^2)^{1/2}$. We measured this quantity
again using all stars within the radius of $2 R_{\rm D}$ in one bin and show the evolution of this bar mode in the upper
panel of Figure~\ref{barmodepatternspeed}. The results confirm the impression from the analysis of the triaxiality
parameter: the bar is initially stronger for stronger interactions during the flyby. There is a subtle difference
however in terms of the time when the first maximum of $A_2$ is reached. We note that the B4 bar grows faster and
reaches its maximum sooner, around $t = 1.3$ Gyr, than the other bars which have the maxima at 1.4, 1.45 and 1.55 Gyr
for B3, B2 and B1, respectively. However, in the later stages of the evolution the hierarchy of the $A_2$ mode is not
preserved. After some decrease in the case of all bars, the one in B2 grows most and ends up highest of all at the end
of the evolution.

\begin{figure}
\centering
\includegraphics[width=7.7cm]{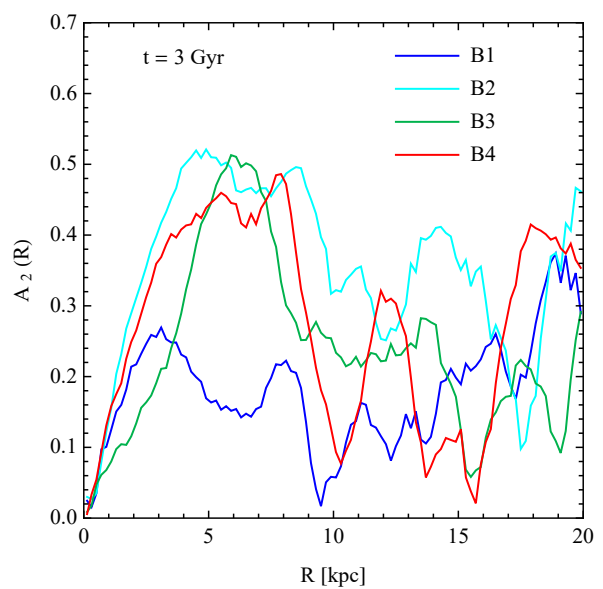}
\caption{Profiles of the bar mode $A_2$ at the end of evolution ($t=3$ Gyr).}
\label{a2profiles}
\end{figure}

\begin{figure}
\centering
\includegraphics[width=4.4cm]{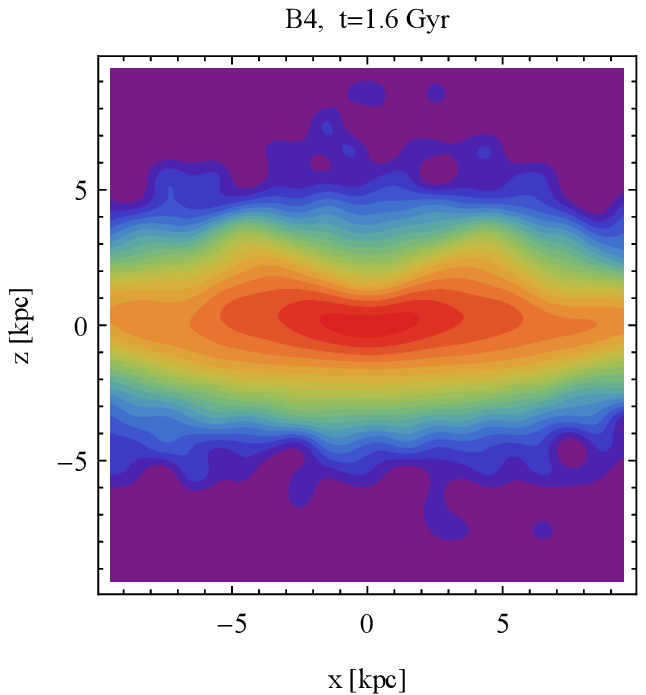}
\includegraphics[width=4.4cm]{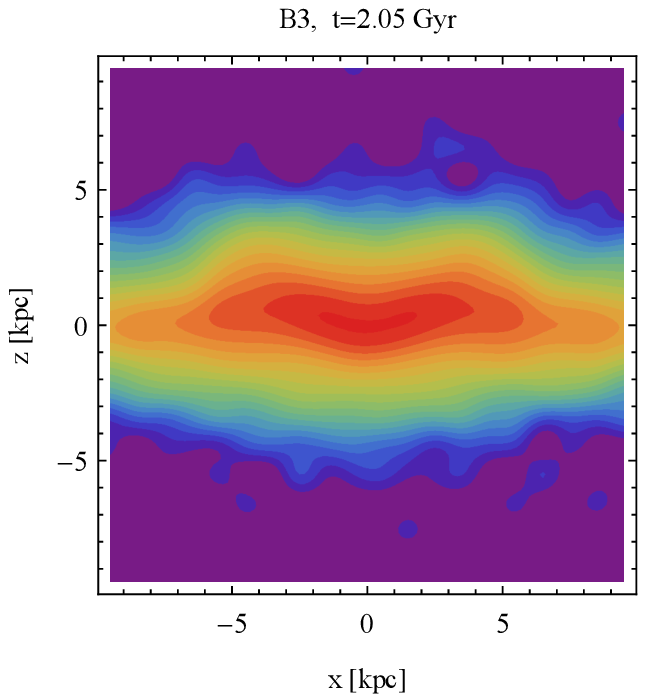}
\includegraphics[width=4.4cm]{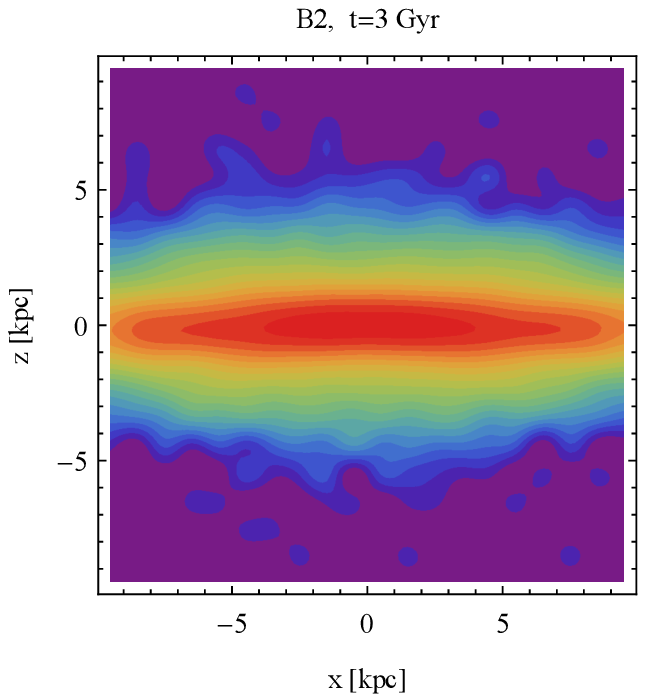}
\includegraphics[width=4.4cm]{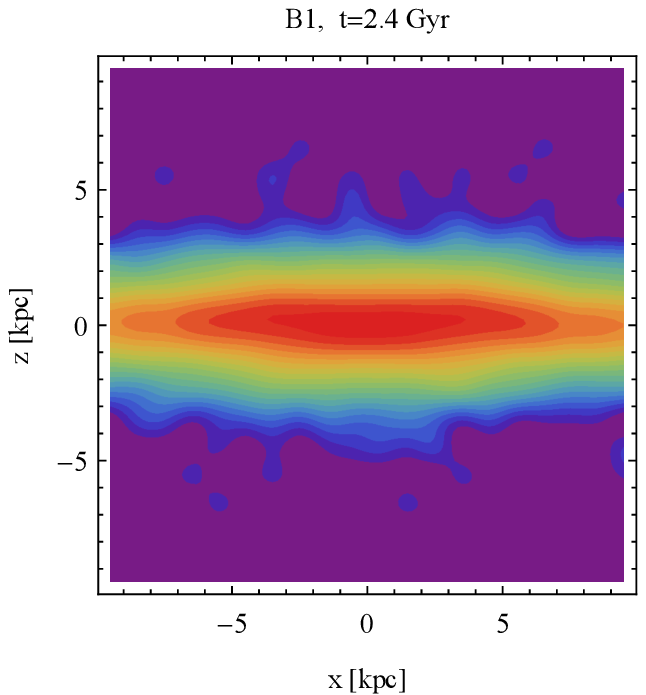} \vspace{0.3cm}
\hspace{0.5cm}
\includegraphics[width=6cm]{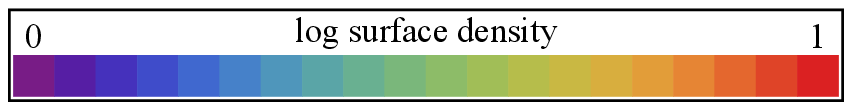}
\caption{Surface density distributions of the stellar component viewed edge-on at the time of maximum distortion out
of the disk plane. For B2 no clear distortion is detected so the final output of the simulation ($t=3$ Gyr) is shown.
The surface density was normalized to the central maximum value in each case and the contours are equally spaced in
$\log \Sigma$ with $\Delta \log \Sigma = 0.05$.}
\label{surden}
\end{figure}

Figure~\ref{a2profiles} shows the profiles of $A_2 (R)$ at the end of the evolution ($t=3$ Gyr). They have a typical
shape strongly increasing at low radii, reaching a maximum at a few kpc and then decreasing. The decrease is not as
smooth as for a typical bar formed in isolation because of additional structures present. The secondary peaks and
variations in the outer radii are due to the presence of the rings and spiral arms also resulting from the interaction.
We note that except for B1, which is rather weak, all the remaining bars are quite strong, with maximum bar mode $A_{\rm
2,max} > 0.4$. We also note that B3 is a bit more spherical in the center, i.e. its $A_2 (R)$ profile grows more slowly
with radius in the inner 5 kpc.

In the lower panel of Figure~\ref{barmodepatternspeed} we plot the evolution of the pattern speed of the bars obtained
by measuring the difference between the orientation of the major axis of the stellar component in two subsequent
simulation outputs. The measurements start at $t=1.1$ Gyr, after the formation of the bar and have been smoothed by
averaging over three subsequent outputs to reduce the noise. The pattern speeds turn out to be rather low, between 10
and 18 km s$^{-1}$ kpc$^{-1}$, characteristic of tidally induced bars and dark matter dominated galaxies. As expected,
the hierarchy of the pattern speeds is inverted with respect to the bar strength as measured by $A_2$, i.e. stronger
bars are slower. Interestingly, this hierarchy is preserved also in the later stages of the evolution, namely the
pattern speed of B2 remains between the values for B1 and B3 in spite of the bar being the strongest in this case.
There is also a dependence between the bar strength and its pattern speed in time for each of the simulated bars: in
general a decreasing trend in the $A_2$ evolution is accompanied by an increasing trend in the pattern speed.
We note that the bars are also slow in terms of the ratio $R_{\rm CR}/a_{\rm b}$ where $R_{\rm CR}$ is the
corotation radius and $a_{\rm b}$ is the bar length. This ratio is of the order of 2 for all our bars \citep{Lokas2018}.

\begin{figure}
\centering
\includegraphics[width=9cm]{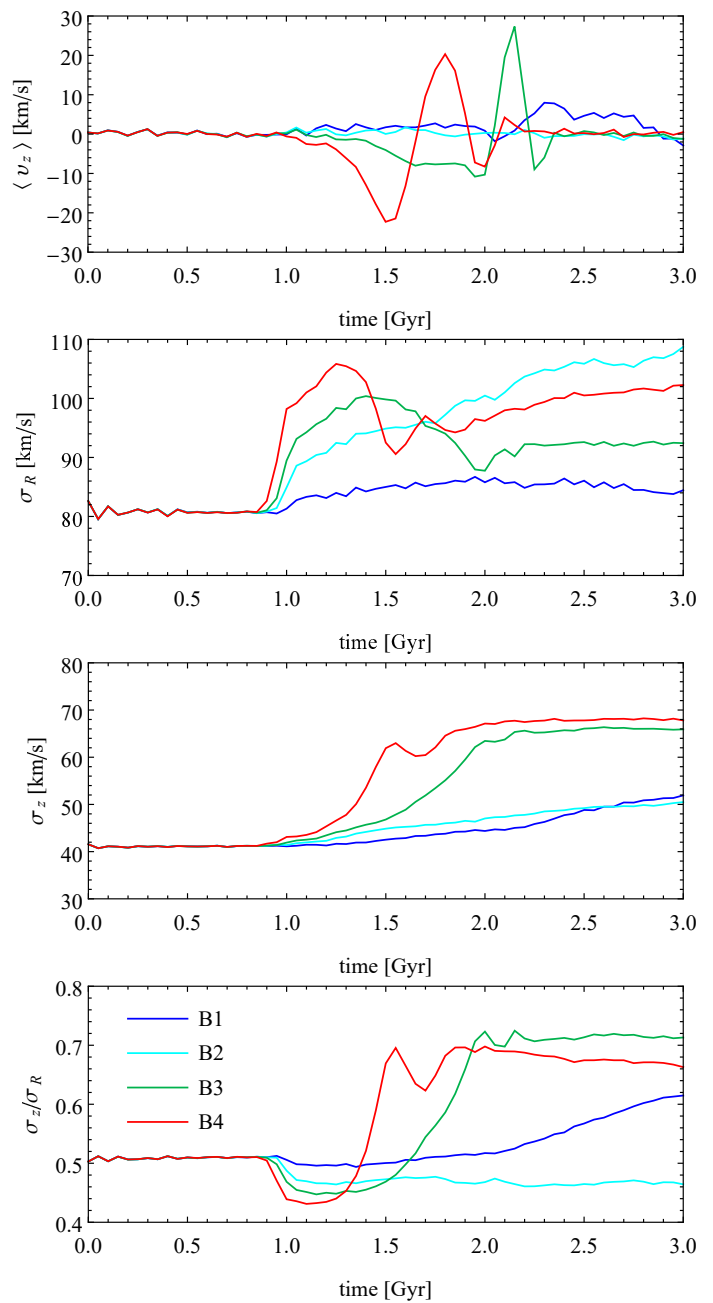}
\caption{Evolution of kinematics of the stellar component in time. The four panels from top to bottom plot the
quantities most relevant for the characterization of buckling: the mean velocity along the vertical axis $\langle \upsilon_z
\rangle$ (first panel), the velocity dispersion along the cylindrical radius $\sigma_R$ (second panel), the velocity
dispersion along the vertical direction $\sigma_z$ (third panel) and the ratio $\sigma_z/\sigma_R$ (fourth panel).
Measurements were made for stars within the radius of $2 R_{\rm D}$.}
\label{kinematics}
\end{figure}

\begin{figure}
\centering
\includegraphics[width=8.9cm]{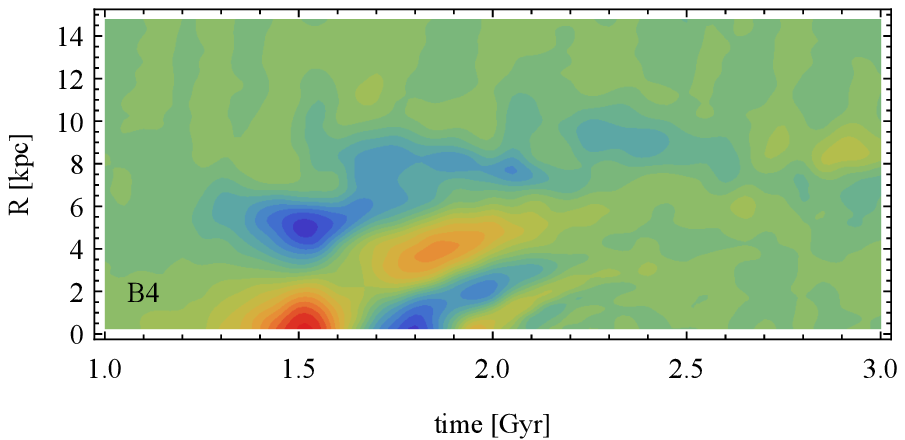}
\includegraphics[width=8.9cm]{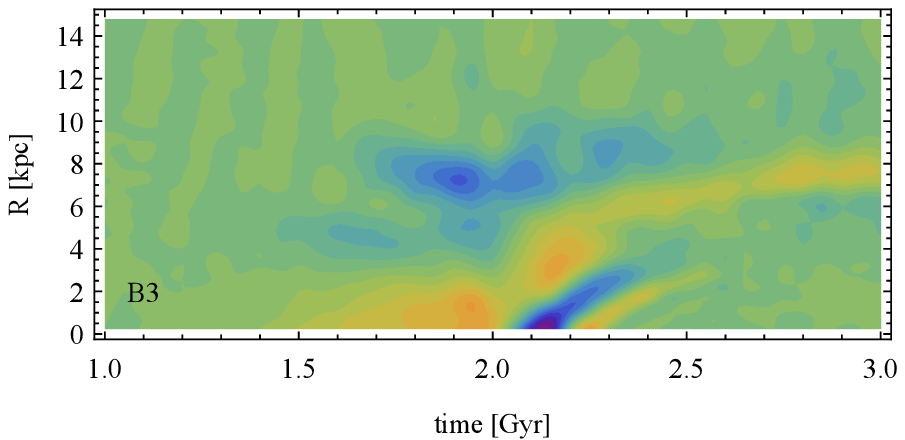}
\includegraphics[width=8.9cm]{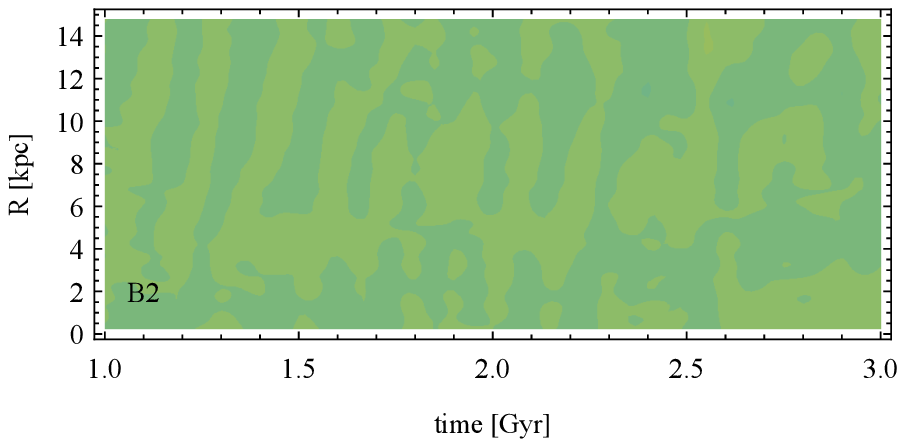}
\includegraphics[width=8.9cm]{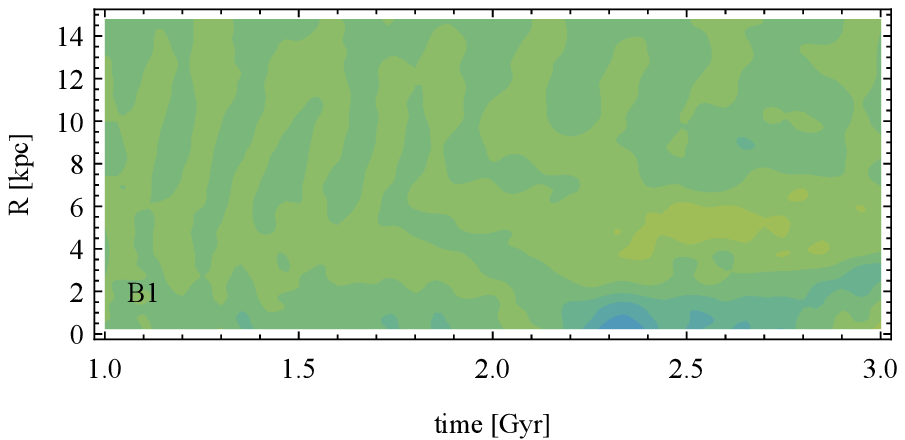}
\vspace{0.5cm}
\hspace{0.7cm} \includegraphics[width=7cm]{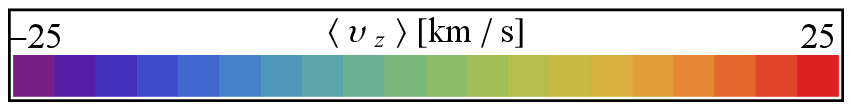}
\caption{Evolution of the profiles of the mean velocity of the stars along the vertical axis,
$\langle \upsilon_z \rangle$, in time for simulations B4-B1 (from top to bottom). Positive velocities point
along the disk's angular momentum vector.}
\label{vzprofile}
\end{figure}

\begin{figure}
\centering
\includegraphics[width=8.9cm]{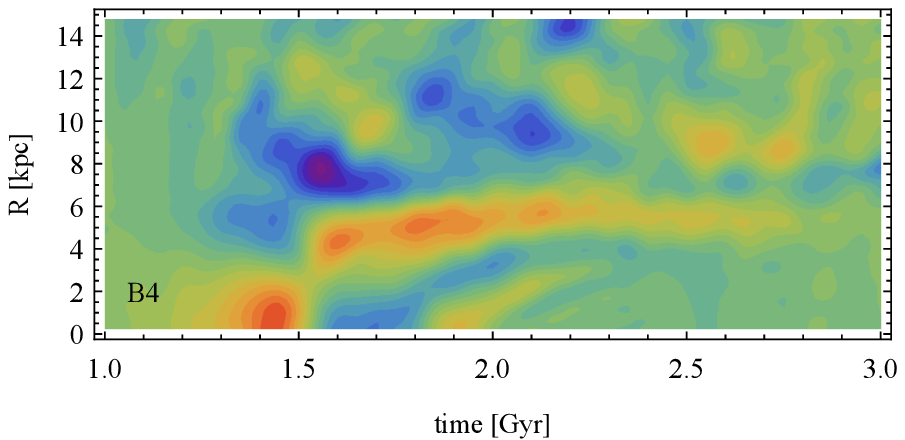}
\includegraphics[width=8.9cm]{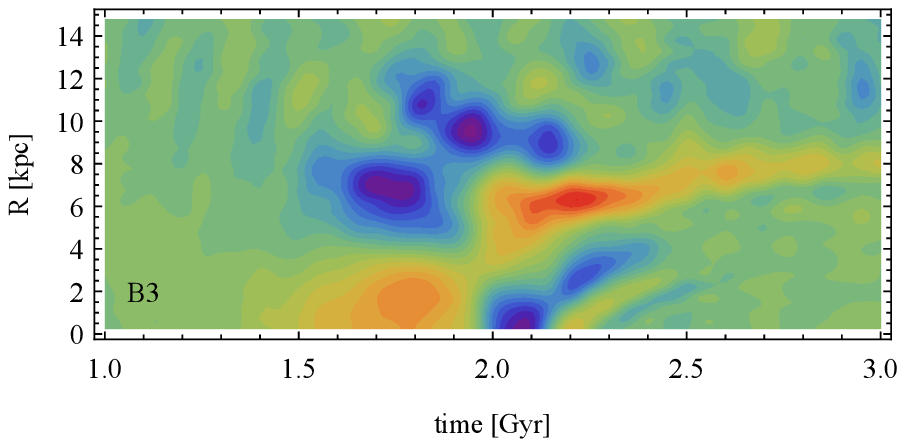}
\includegraphics[width=8.9cm]{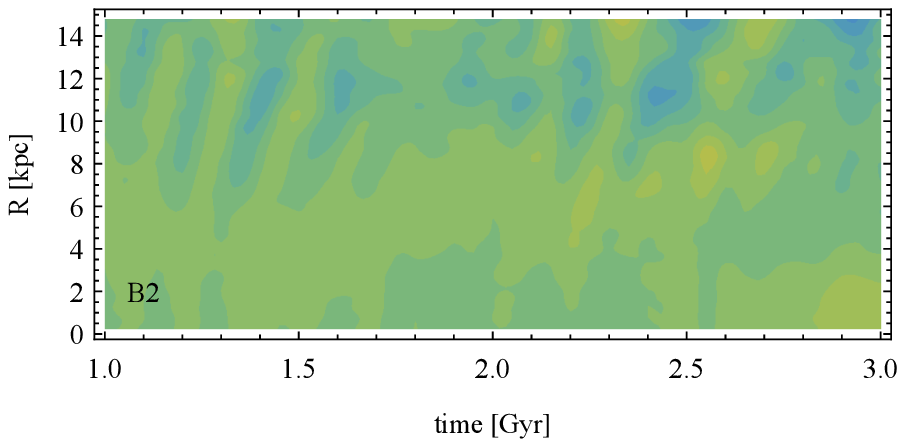}
\includegraphics[width=8.9cm]{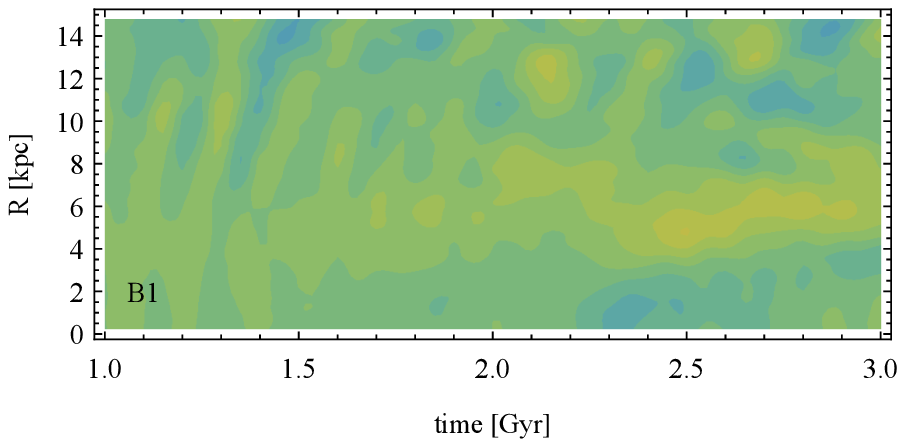}
\vspace{0.5cm}
\hspace{0.7cm} \includegraphics[width=7cm]{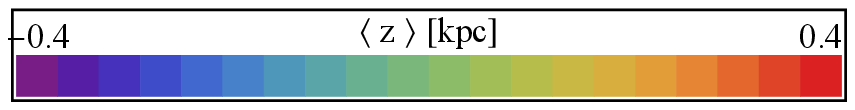}
\caption{Evolution of the profiles of the mean distortion of the positions of the stars
along the vertical axis, $\langle z \rangle$, in time for simulations B4-B1 (from top to bottom).
Positive distortions are upwards, along the disk's angular momentum vector.}
\label{meanzprofile}
\end{figure}

\begin{figure}
\centering
\includegraphics[width=4.4cm]{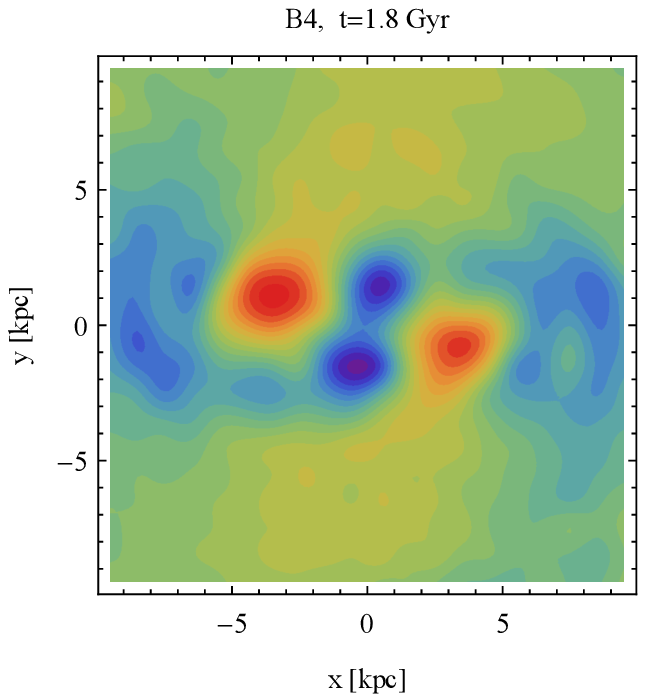}
\includegraphics[width=4.4cm]{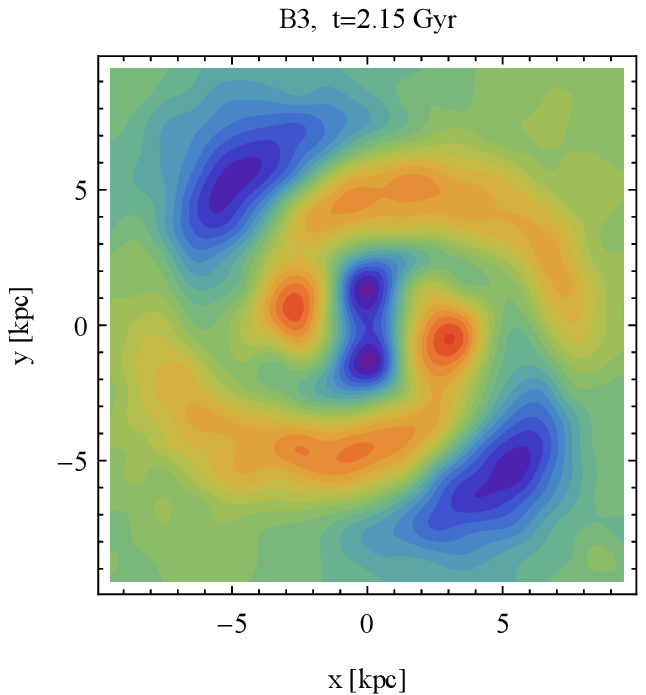}
\includegraphics[width=4.4cm]{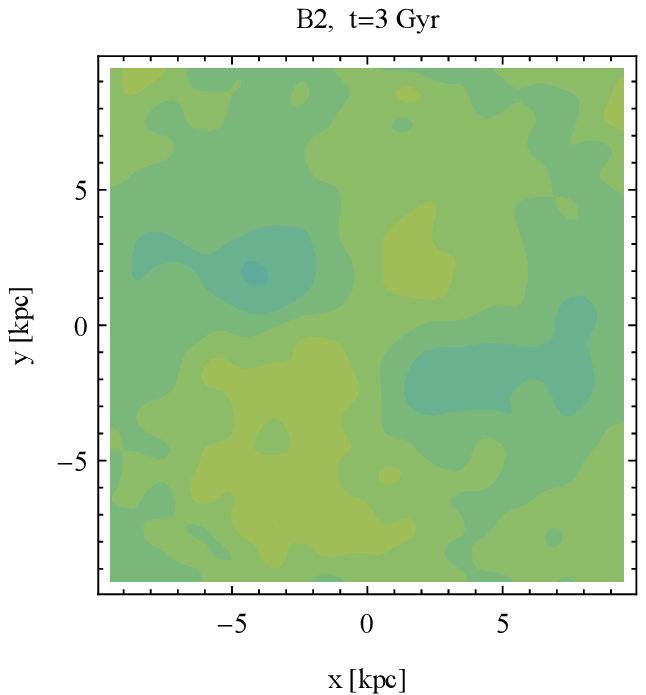}
\includegraphics[width=4.4cm]{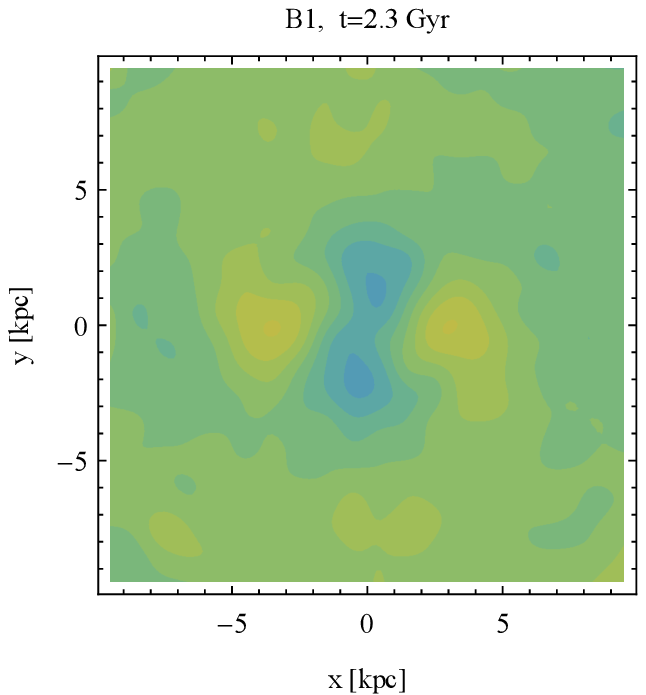} \vspace{0.3cm}
\hspace{0.5cm}
\includegraphics[width=6cm]{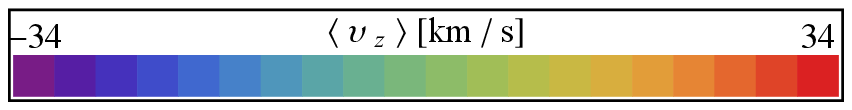}
\caption{Face-on maps of the mean velocity of the stars along the vertical axis, $\langle \upsilon_z \rangle$,
for simulations B4-B1 at the time of strongest buckling, except for B2 where the final output
of the simulation ($t=3$ Gyr) is shown. Positive velocities point along the disk's angular momentum vector.}
\label{vzmap}
\end{figure}

\section{Buckling instability}

The decrease of the strength of the bar mode soon after the formation of the bar, as seen in the upper panel of
Figure~\ref{barmodepatternspeed}, is typically a signature of buckling instability which tends to weaken the bar. We
verify that this phenomenon indeed occurs in our bars by visual inspection of the surface density distribution of the
stars in the edge-on view, i.e. along the intermediate axis of the bar. A few examples of such distributions are shown
in Figure~\ref{surden}, one for each simulation. We see that strong distortions of the bar out of the disk plane
are indeed present, especially in the case of B4 and B3, while only a weak distortion is present for B1 and
almost none for B2. As discussed in more detail below, the output times used in Figure~\ref{surden} were
selected as those when the close to maximum distortion is seen for a given simulation, except for B2 where we took the
final output ($t=3$ Gyr).

In order to further explore the phenomenon and determine its timescale, we look at the kinematic evolution of the bars.
The distortions must involve systematic departures out of the disk plane of significant numbers of stars and these may
manifest themselves in ordered motion along the $z$ axis. We measure this by calculating the mean velocity of stars
along $z$,  $\langle \upsilon_z \rangle$, as a function of time using stars within $2 R_{\rm D}$. This quantity is
shown in the upper panel of Figure~\ref{kinematics} measured with respect to the velocity of the center of the
galaxy, estimated iteratively down to the radius of 0.5 kpc. The velocities were assumed to have a positive sign if
they pointed along the positive $z$ axis, which was chosen to point in the same direction as the galaxy's angular
momentum vector. Clearly, a strong signal of mean velocity out of the disk plane is present, especially in the case of
B4 and B3, reaching $20-30$ km s$^{-1}$. Interestingly, a weak signal, at the level of 8 km s$^{-1}$, is also visible
for the weakest bar B1, formed in the weakest interaction, but not in the intermediate case B2, where the mean velocity
does not exceed the noise. The kinematic signal occurs earlier for stronger bars, first for B4, then for B3 and at the
latest time for B1. We note that the streaming motion with respect to the center in cases B3 and B4 has a
negative sign first which changes to positive later on. This means that the stars move downwards first, creating a
frown-like distortion of the bar, and upwards later, changing into a smile-like distortion. This is confirmed by
following the evolution of the edge-on images like those in Figure~\ref{surden} in time.

Buckling is known to increase the velocity dispersion along the $z$ axis and the thickness of the bar. In the
second and third panel of Figure~\ref{kinematics} we plot the evolution of the velocity dispersions in the radial and
vertical direction, $\sigma_R$ and $\sigma_z$, respectively. While $\sigma_R$ is a measure of the bar strength, because
it reflects the amount of radial motion in the bar, $\sigma_z$ quantifies the effect of buckling in increasing the
amount of vertical random motion in the bar which is related to its thickness. Indeed, the evolution of $\sigma_R$
reflects that of the bar mode $A_2$ in the upper panel of Figure~\ref{barmodepatternspeed}. On the other hand, the
evolution of $\sigma_z$ correlates with the velocity signature of buckling in the upper panel of
Figure~\ref{kinematics}. Its increase is strongest in the case of B4 and B3, much weaker for B1 and occurs sooner for
bars buckling earlier. Similar increase is seen in the thickness of the stellar component as measured by the axis ratio
$c/a$ plotted in the middle panel of Figure~\ref{shape}.

In the lower panel of Figure~\ref{kinematics} we show the evolution of the ratio
$\sigma_z/\sigma_R$ with time. We see that right after the formation of the bar, the ratio drops and it does so more
strongly for stronger bars. The more the ratio drops, the faster the buckling seems to occur, and the sooner the
ratio $\sigma_z/\sigma_R$ grows at least for B4 and B3. This dependence is however broken for the weaker bars, B2 and
B1: although the drop in B1 is smallest, this bar does buckle weakly later on, while B2, which shows a bigger drop, does
not buckle at all and its $\sigma_z/\sigma_R$ remains at the same low level until the end of evolution. Let us
note that in the case of B2 both $\sigma_z$ and $\sigma_R$ increase at a similar rate in time, i.e. there is an
approximately linear growth of $\sigma_z$ in time, probably due to heating, but no speed up of this growth takes place
as it does in the case of B1.

In order to obtain further insight into the kinematic structure of the bar at the time of buckling, in
Figure~\ref{vzprofile} we plot the maps showing the evolution of the profiles $\langle \upsilon_z \rangle (R)$ in time.
While for the global, single-value, measurement of $\langle \upsilon_z \rangle (<2 R_{\rm D})$ shown in the upper panel
of Figure~\ref{kinematics} we used the reference frame centered on the very center of the galaxy, here we measure the
velocities along $z$ with respect to the average, bulk motion of the galaxy estimated as the mean velocity of all stars
within the radius of 10 kpc. As a result, the maps in Figure~\ref{vzprofile} show the motion of the center going up and
the outer parts going down first (around $t=1.5$ Gyr for B4 and $t=1.95$ Gyr for B3), corresponding to the frown-like
phase detected in the global measurement of Figure~\ref{kinematics}. Later on, in the second phase, the central parts
of the bar move down and the outer parts move up and the distortion is smile-like. After these two main phases the bars
start to oscillate even more, i.e. there are more changes in the sign of the velocity along the radius $R$, until no
large scale ordered motion is present any more.

In Figure~\ref{meanzprofile} we plot the profiles of the mean distortion of the stellar component along $z$,
$\langle z \rangle (R)$, as a function of time. As for the velocities, the measurements were done with respect to the
mean position of the stars within the radius of 10 kpc. Also in this case, the first phase of frown-like distortion is
seen, followed by a smile-like phase and more complicated patterns later on. As expected, the maximal distortions are
shifted with respect to the maximal velocities along $z$, for example the maximum velocity at $t=1.5$ Gyr for B4
corresponds to the change in the sign of the distortion at this time. Although the streaming velocities and distortions
are much weaker in B1 than in B4 and B3, the characteristic pattern is still present. In the case of B2, some
distortion in the outer radii are seen, but no corresponding signal is present in the velocity map, confirming our
earlier claim that the bar in B2, in spite of being quite strong, does not buckle.

While the evolutionary plots of Figures~\ref{vzprofile} and \ref{meanzprofile} indeed provide more information
on the character of the distortions, they still simplify the picture a little because the measurements are done in
radial bins thus averaging the dependence of the data on the orientation. A fuller picture can be obtained by looking
at the face-on maps of the mean vertical velocity distribution, as shown in Figure~\ref{vzmap}. Here the velocities
were binned into 1 kpc $\times$ 1 kpc bins and the measurements were done again with respect to the mean velocity of the
stars within 10 kpc. The examples shown in the Figure correspond to the second phase of buckling, when the velocity is
maximal and in the smile-like direction, except for B2 where we show the final output. In each panel the bar is
oriented along the $x$ axis ($y=0$) and the galaxy is rotating anti-clockwise. An interesting pattern is revealed: the
two regions of the dominant upward motion (in red) are located along the bar, while another pair of two regions moving
downwards (in blue) are closer to the center and oriented perpendicular to the bar. More structure is present further
out, related to the spiral arms which are particularly well visible in the case of B3. We note that a similar, although
weaker, pattern of two regions moving upward and two moving downward is seen in B1, while no such structure is present
in B2. While there is some motion in B2, it is on a larger scale and not related to the bar.

\begin{figure}
\centering
\includegraphics[width=7.7cm]{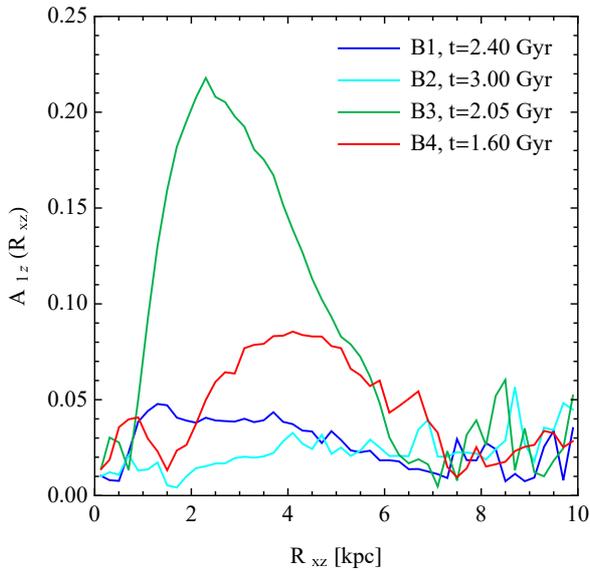}
\caption{Profiles of the asymmetry measure $A_{1z}$ of the stellar component in the edge-on view during buckling.}
\label{a1profiles}
\end{figure}

To complete our description of the distortions due to buckling and facilitate a comparison with other studies we
calculated another commonly used measure of the asymmetry in the form of the $m=1$ mode of the Fourier decomposition of
the surface distribution of stars in the edge-on view (projected along the intermediate $y$ axis): $A_{mz} (R_{xz}) = |
\Sigma_j \exp(i m \theta_j) |/N_s$. This is similar to the standard mode calculation, where $\theta_j$ is the azimuthal
angle of the $j$th star and the sum is up to the total number of $N_s$ stars, but now the radius $R_{xz}$ is measured
in the $xz$ plane, $R_{xz} = (x^2 + z^2)^{1/2}$. Examples of the $A_{1z} (R_{xz})$ profiles for the times where the
maximum distortions occur (according to Figure~\ref{meanzprofile} for the second phase of buckling) for different
simulations are shown in Figure~\ref{a1profiles}.

\begin{figure}
\centering
\includegraphics[width=7.4cm]{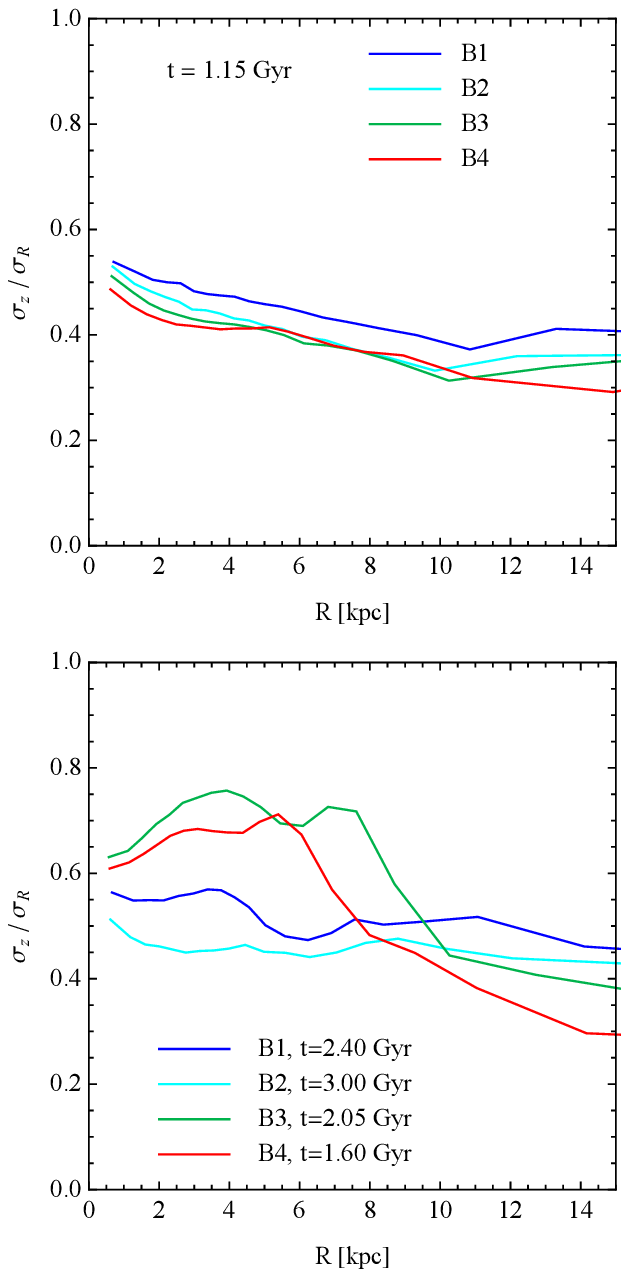}
\caption{Profiles of the dispersion ratio $\sigma_z/\sigma_R$ of the stellar component. The upper panel shows the
profiles just after the formation of the bars but before the onset of buckling. The lower panel shows
the ratio at the time of buckling.}
\label{bucklingdispersions}
\end{figure}

\begin{figure*}
\centering
\includegraphics[width=5.5cm]{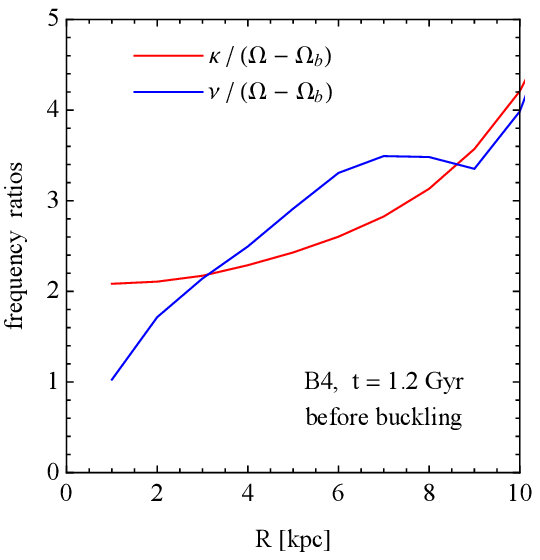}
\includegraphics[width=5.5cm]{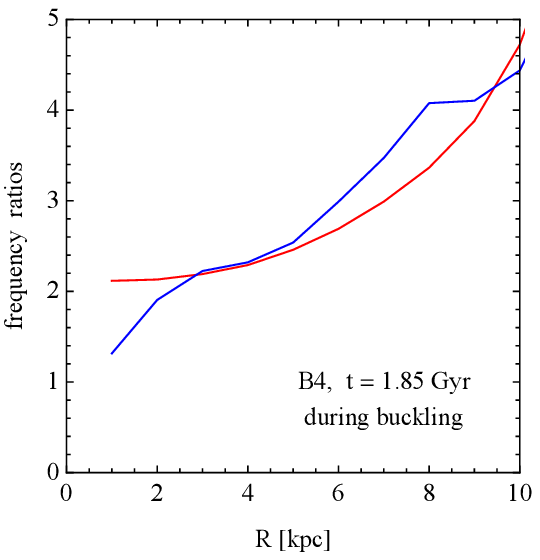}
\includegraphics[width=5.5cm]{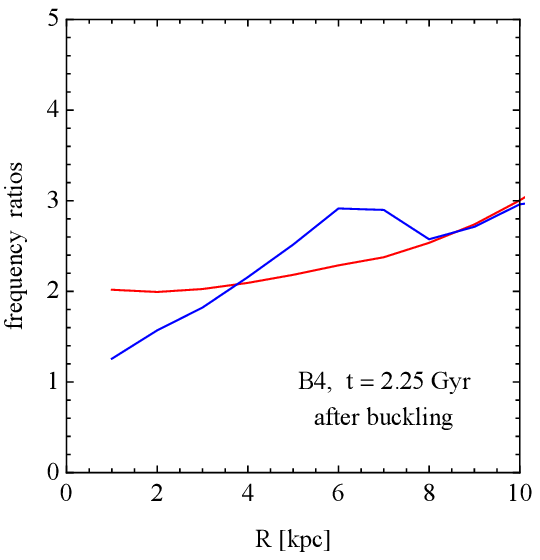}
\caption{Frequency ratios $\kappa/(\Omega-\Omega_{\rm p})$ and $\nu/(\Omega-\Omega_{\rm p})$ before (left panel),
during (middle panel) and after buckling (right panel) for the bar formed in simulation B4.}
\label{freqratios}
\end{figure*}

For the strongly buckling bars B3 and B4 the maxima of $A_{1z} (R_{xz})$ are 0.22 and 0.09 respectively, but the
profiles peak at different radii: in the case of B3 the distortion is strongest closer to the center of the galaxy,
around $R_{xz} = 2$ kpc while for B4 it occurs further out, at $R_{xz} = 4$ kpc. The times of this maximum distortion
are also different. In the case of B4 the maximum occurs at $t = 1.6$ Gyr after the start of the simulation, while for
B3 the maximum distortion is present much later, at $t=2.05$ Gyr. Note that the hierarchy of the maxima in B4
and B3 would we reversed had we considered the first phase of buckling: the distortion is then stronger for B4. For B1
the values of $A_{1z} (R_{xz})$ are much lower reaching 0.05 near the center of the galaxy with a slight decreasing
trend with radius, and the maximum occurs around $t = 2.4$ Gyr. For B2 they are low at all times, with the radial trend
rather opposite, so we show the example at the end of the evolution, $t = 3$ Gyr. The times when the maximal
distortions occur for each simulation were the same as those for which we show the edge-on galaxy images in
Figure~\ref{surden}.

\section{Discussion}

We studied the evolution of bars formed tidally as a result of flyby interactions of different strength, in particular
the buckling instability that occurs in them. We find that strong buckling takes place in stronger bars, although the
weakest bar in our sample also buckles a little. We found no direct relation between the value of the ratio of velocity
dispersions in the vertical and radial direction, $\sigma_z/\sigma_R$, and the bar's susceptibility to buckling. While
our weakest bar B1 with highest $\sigma_z/\sigma_R$ does buckle, a stronger one with lower $\sigma_z/\sigma_R$ present
in simulation B2 does not. This suggests that the nature of buckling instability is not related to the fire-hose
instability known from plasma physics.

In order to further explore the dependence of buckling on the ratio $\sigma_z/\sigma_R$ after bar formation, we plot in
the upper panel of Figure~\ref{bucklingdispersions} the profiles of this ratio for all our simulations at the time right
after the formation of the bars ($t=1.15$ Gyr). We can see that stronger bars have this ratio systematically lower
within $2 R_{\rm D}$. In particular, the ratio has a significantly larger value for our weakest bar B1. If buckling
was strictly related to the value of the ratio $\sigma_z/\sigma_R$ then B1 should be the least prone to the
instability, which is not the case. Instead, the bar which does not buckle is the one in simulation B2 which has
the $\sigma_z/\sigma_R$ profile very similar to B4 and B3 which buckle strongly.
In the lower panel of Figure~\ref{bucklingdispersions} we plot the same quantity, $\sigma_z/\sigma_R$, but at the time
of the second phase of buckling, when the distortion of the bars out of the disk plane in largest. We see that the ratio
$\sigma_z/\sigma_R$ is now strongly increased in most cases as a result of buckling, but remains low for B2 which
did not buckle. While the values of $\sigma_z/\sigma_R$ increased a little for this simulation in comparison
with the earlier time (shown in the upper panel of the Figure), it was probably due to heating and not buckling since
no strong vertical streaming motions associated with the bar were found in our analysis of this case in the previous
section.

These results therefore support the alternative interpretation of the nature of buckling instability which states that
it results from orbital instabilities and orbit trapping in the vicinity of the horizontal and vertical inner Lindblad
resonances. To check if this interpretation is plausible we calculated the ratios $\kappa/(\Omega-\Omega_{\rm p})$ and
$\nu/(\Omega-\Omega_{\rm p})$, where $\kappa$, $\Omega$ and $\nu$ are the radial, circular and vertical frequencies and
$\Omega_{\rm p}$ is the pattern speed of the bar shown in Figure~\ref{barmodepatternspeed}. For this purpose we follow
\citet{Pfenniger1990} and \citet{Pfenniger1991} and use an axisymmetric approximation and the Poisson equation in the
form $\nu^2 = 4 \pi G \rho + 2 \Omega^2 - \kappa^2$ which we evaluate at $z=0$ expressing the density by the surface
density measured at each simulation output.

Examples of the profiles of frequency ratios $\kappa/(\Omega-\Omega_{\rm p})$ and $\nu/(\Omega-\Omega_{\rm p})$ for
our strongest bar B4 are shown in Figure~\ref{freqratios}. As discussed by \citet{Pfenniger1990}, a resonance zone where
$\kappa/\nu \approx 1$ is always present in any potential which goes from spherical symmetry ($\kappa/\nu =2$) to a flat
disk ($\kappa/\nu \rightarrow 0$). This is what we see in the left panel of Figure~\ref{freqratios} along the
cylindrical radius $R$ which shows the frequency ratios before the bar buckles. Clearly, the resonance occurs only at
one radius and is therefore very narrow and not effective in modifying the orbits of stars. However, following the
evolution of the ratios to later times we find that at the time of buckling (middle panel of Figure~\ref{freqratios})
the frequency ratios coincide over a significant range of radii, 3 kpc $< R <$ 5 kpc, and are both close to the value of
2 characteristic of inner Lindblad resonances. At this time the resonance is thus much wider and able to dispatch stars
out of the disk plane. After buckling (right panel of Figure~\ref{freqratios}) the resonance becomes narrow again in
this radial range.

Similar behavior is found in simulations B3 and B1, although for B4 it looks most convincing. For B2 $\kappa$
and $\nu$ coincide in a narrow range of radii and times but the buckling does not lift off. The reason why B2 does not
buckle may be related to the presence of spiral arms. While the tidally induced bars studied here are particularly
useful because they originate from the same initial disky galaxy, the difficulty such configurations entail is the
formation of tidally induced spiral arms in addition to bars. The spiral arms are quite strong in the case of B2, even
at the end of evolution, as confirmed by the presence of high secondary peaks of the $A_2$ profile at larger radii in
Figure~\ref{a2profiles}. After their formation, the spiral arms wind up and may disturb the orbital structure of the
bar. In the case of B4 and B3 the bars may be strong enough to remain unaffected and in the case of B1 the spiral arms
may be too weak, but for B2 they may affect the bar. However, even if this is the case, this circumstance actually
speaks in favor of our interpretation of buckling in terms of vertical orbital instabilities. While the orbital
structure of B2 may have been disturbed by the spiral arms its $\sigma_z/\sigma_R$ remained low and should have led to
buckling if this was the quantity that controlled the occurrence of this phenomenon.

In summary, although to some extent similar to the fire-hose instability, the buckling seen in the tidally induced bars
studied here does not relate directly to the ratio $\sigma_z/\sigma_R$, as would be required in this case. We conclude
in agreement with \citet{Pfenniger1991} and \citet{Pfenniger1996} that the phenomenon of buckling is probably due to
vertical instability of stellar orbits supporting the bar.

\begin{acknowledgements}
      This work was supported in part by the Polish National Science Center under grant 2013/10/A/ST9/00023.
Insightful comments from the anonymous referee are kindly appreciated.
\end{acknowledgements}

\end{document}